# Adaptive Priority-Based Downlink Scheduling for WiMAX Networks

Shih-Jung Wu, Shih-Yi Huang , and Kuo-Feng Huang

*Abstract:* **Supporting quality of service (QoS) guarantees for diverse multimedia services are the primary concerns for WiMAX (IEEE 802.16) networks. A scheduling scheme that satisfies QoS requirements has become more important for wireless communications. We propose a downlink scheduling scheme called adaptive priority-based downlink scheduling (APDS) for providing QoS guarantees in IEEE 802.16 networks. APDS comprises two major components: priority assignment and resource allocation. Different service-type connections primarily depend on their QoS requirements to adjust priority assignments and dispatch bandwidth resources dynamically. We consider both starvation avoidance and resource management. Simulation results show that our APDS methodology outperforms the representative scheduling approaches in QoS satisfaction and maintains fairness in starvation prevention.**

## I. INTRODUCTION

In recent years, wireless broadband networks have undergone significant development. As evidenced by current research interests, next generation wireless broadband networks will combine both wireless and broadband networks [1, 2]. The maturation of wireless technologies and development of Internet services have led to an increase in demand for wireless multimedia transmission, data communication, and other mobile services. Many researchers consider the wireless metropolitan area network as a potential solution for mobile communication technology issues. The IEEE 802.16 standard on broadband wireless access is the main technology standard for developing wireless broadband network systems. The purpose of IEEE 802.16 is to build high coverage and high transmission rate wireless metropolitan area networks that provide quality "last mile" wireless access [7–9]. The media access control (MAC) sub-layer in the data link layer of an open system interconnection architecture manages both media and resource utilization. As network performance is significantly affected by the quality of the scheduling algorithm, the design of our scheduler in the MAC layer was a major focal point [5, 12]. In addition, scheduling algorithms for wireless networks are much more complex than those for wired networks because of channel quality variations and radio resource limits. The objective of IEEE 802.16 is to provide highly stable wireless access networks with high transmission rates and QoS [3, 4]. The scheduling algorithms can be categorized into two types: service-based and connection-based. For service-based algorithms, scheduling is determined according to service type. For connection-based algorithms, all of the connections are scheduled as the same service type [13–33]. In this paper, we propose an adaptive priority-based downlink scheduling (APDS) algorithm to improve network performance. The algorithm makes dynamic adjustments to bandwidth allocation according to user demand. Moreover, a weight-based proportional fairness scheme has been proposed to decrease starvation of lower level services (i.e., best effort services). The rest of the paper is organized as follows: section 2 presents the scheduling and bandwidth allocation scheme in the proposed APDS; section 3 presents the results of our simulations; and section 4 presents our conclusions.

## II. ADAPTIVE PRIORITY-BASED DOWNLINK SCHEDULING SCHEMES

In general, there are three components in an IEEE 802.16 network: the base station (BS), subscriber station (SS), and mobile station (MS). SS and MS are both clients; the difference is that MS clients are mobile. The objective of this paper is to describe a downlink scheduling scheme that exhibits better performance for the IEEE 802.16 standard. IEEE 802.16 is a connection-oriented wireless communication technology. Each connection in an IEEE 802.16 network is identified by a unique connection ID (CID) that is assigned by the BS. The connection provides bandwidth resources on a downlink or uplink connection. We dynamically adjust the bandwidth allocation for downlinks with downloads to meet QoS restrictions. For each user, we guarantee service quality according to his QoS parameter and avoid starving lower service levels. With these goals, we propose a downlink scheduling scheme called Adaptive Priority-based Downlink Scheduling (APDS). APDS operates in a point-to-multipoint network architecture with time division duplexing technology for data transmission. The proposed algorithm is a service-based centralized scheduling algorithm. It is also a non-work-conserving scheduling algorithm because the scheduling is performed before each frame [5, 35]. Admission control [36, 37] is not a main consideration in this paper.

### A. SYSTEM ARCHITECTURE

In the proposed scheme, each connection will be assigned a priority that identifies the transmission order. APDS improves the QoS guarantee by dynamically adjusting priorities while also taking QoS restrictions into consideration. The QoS parameters defined by the 802.16 standard will be considered and

Shih-Jung Wu is with the Department of Innovative Information and Technology, Tamkang University in Taiwan (R.O.C.)., email: wushihjung@mail.tku.edu.tw.

Shih-Yi Huang is with the Chunghwa Telecom Laboratories in Taiwan (R.O.C.) , email: shiyi1122@hotmail.com.

Kuo-Feng Huang is with Department of Information Technology and Mobile Communication, Taipei College of Maritime Technology in Taiwan (R.O.C.), 896410148@s96.tku.edu.tw.

quantified to allow the scheduler's adjustments to be more flexible and precise [4, 6]. Table 1 shows the QoS parameter definitions used in this paper.

Table 1. DEFINITIONS OF THE QOS PARAMETERS

| Notation | Definition |
|---|---|
| $\Gamma_i$ | Maximum sustained traffic rate of connection $i$ |
| $\gamma_i$ | Minimum reserved traffic rate of connection $i$ |
| $\zeta_i$ | Maximum latency of connection $i$ |
| $\varsigma_i$ | Unsolicited grant interval of connection $i$ |
| $\delta_i$ | Tolerated jitter of connection $i$ |

The five service types defined by IEEE 802.16 are categorized as two services: delay-constrained services (DCS) and throughput-guaranteed services (TGS). DCSs include UGS (unsolicited grant service), ERT-VR (extended real-time variable), and RT-VR (real-time variable rate). TGSs include NRT-VR (non-real-time variable rate) and BE (best effort). With APDS, DCS requests are satisfied before TGS requests. Furthermore, the connection with the lowest priority in the same category promotes its own priority to avoid starvation or connection breakdown. As shown in Fig. 1, for DCSs, an RT-VR can be promoted to an ERT-VR, and the ERT-VR can then be promoted to a UGS. For TGSs, a BE can be promoted to an NRT-VR.

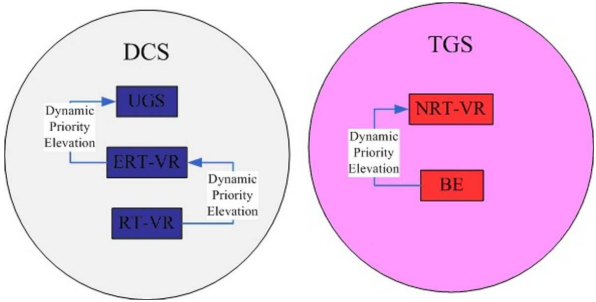

Fig. 1. Dynamic priority elevation.

There are two phases included in APDS: priority assignment and resource allocation. As shown in Fig 2, priority assignment and resource allocation can be divided into two separate operations. The priority assignment phase comprises both connection rankings and priority elevations. Connection ranking determines the priority of connections by their specified parameters. Priority elevation avoids starvation and connection breakdowns by promoting the connection with the lowest priority. For the resource allocation phase, quantification and allocation of bandwidth requirements are performed. Bandwidth requirement quantification calculates the upper and lower bounds of possible bandwidth requests for each connection, allowing dynamic bandwidth allocation by aggregating the upper and lower bounds of all bandwidth requests. Bandwidth requirement allocation allocates bandwidth according to the connection ranking.

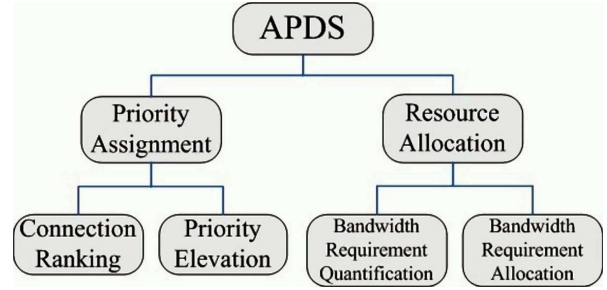

Fig. 2. APDS architecture.

### B. Priority Assignment

As shown in Fig. 3 two types of operations are connection rankings and priority elevations. For connection ranking, we consider two factors: emergent degree and satisfactory degree. The current average latency is calculated as the emergent degree because of the strict latency requests for DCSs. For TGSs, the allocated bandwidth in the last frame is considered as the satisfactory degree for ranking all connections. Moreover, five ranking queues for different service types are used to store ranked connections. To satisfy the QoS requests and avoid lower priority connection breakdowns in the APDS, we implemented an emergent queue for DCSs and a service interrupt counter for TGSs.

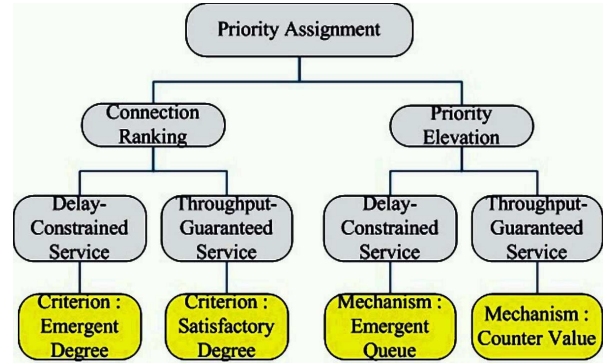

Fig. 3. Priority assignment architecture.

### B.1 Connection Ranking

#### B.1.1 DCS

We use the symbols $RQ_{UGS}^{DL}$, $RQ_{ERT}^{DL}$ and $RQ_{RT}^{DL}$ to identify the downlink ranking queues for UGs, ERT-VRs, and RT-VRs. The emergent degree is used to elevate ERT-VR and RT-VR services. For the downlink, an emergent degree is calculated using the tolerable latency (determined by the average remaining wait time). All DCS, UGS, ERT-VR and RT-VR services are viewed as variable bit rate (VBR) for the downlink [4, 6]. Here are select parameters that are used in the following algorithms: $T_{frame}$ represents the length of a frame, $\zeta_i$ represents the maximum latency of connection $i$, and $N_i$ represents the packet size in connection $i$. $T_i^a(j)$ is the arrival time of the transferred packet at the MAC layer. $T_i^w(j)$ is the wait time of each packet in the MAC layer. This wait time can be calculated using equation 1. $T_i^g(j)$ is the guard time (tolerable wait time) of each packet, as shown in equation 2. $T_c$ is the current system time. $\zeta_i$ is the maximum latency of connection $i$.

$$T_i^w(j) = T_c - T_i^a(j) \quad (1)$$

$$T_i^g(j) = \zeta_i - T_i^w(j) \quad (2)$$

$L_i$ is the tolerable latency of connection $i$ and identifies the emergent degree of connection $i$. This latency is calculated by equation 3. We calculate $Li$ to normalize algorithms 1, 2, and 3. Then, we arrange the order according to this normalized value for every connection.

$$L_i = \sum_{j=1}^{N_i} \frac{T_i^g(j)}{N_j} \quad (3)$$

-UGS: In Algorithm 1, $N_{UGS}^{DL}$ represents the number of UGSs in the downlink, and $\Omega_{UGS\_DL}^{CID}$ represents the CID set of the UGSs in the downlink. The ranking is determined by the emergent degree and then sequentially pushed onto the ranking queues, $RQ_{UGS}^{DL}$.

---

**Algorithm 1** Priority assignment for a UGS connection

**Require:** $N_{UGS}^{DL}$,
$\quad \Omega_{ERT\_DL}^{CID} = \{CID_1 \ldots CID_i\}, \forall = 1 \ldots N_{UGS}^{DL}$

**Ensure:** $RQ_{UGS}^{DL}$

---

1: BEGIN
2: **for** $i = 1$ to $N_{UGS}^{DL}$ **do**
3: $\quad L_i = \sum_{j=1}^{N_j} \frac{T_i^g(j)}{N_j}, \forall = 1 \ldots N_{UGS}^{DL}, \forall = j \ldots N_j$
4: $\quad L_max = argmax L_i, \forall = 1 \ldots N_{UGS}^{DL}$
5: $\quad L_i = 1 - \frac{L_i}{L_{max}}, \forall = 1 \ldots N_{UGS}^{DL}$
6: $\quad$ /* $P_j$ : the $j$th packet for connection i */
7: **end for**
8: **while** $\Omega_{UGS\_DL}^{CID} > 0$ **do**
9: $\quad CID_{max} = argmin_i L_i$
10: $\quad CID_{max}$ add to $RQ_{UGS}^{DL}$
11: $\quad \Omega_{UGS\_DL}^{CID} = \Omega_{UGS\_DL}^{CID} - CID_{max}$
12: **end while**
13: END

---

-ERT-VR: In Algorithm 2, $N_{ERT}^{DL}$ represents the number of ERT-VRs in the downlink, and $\Omega_{ERT\_DL}^{CID}$ represents the CID set of the ERT-VRs in the downlink. The ranking is determined by the emergent degree and then sequentially pushed onto the ranking queues, $RQ_{ERT}^{DL}$.

---

**Algorithm 2** Priority assignment for an ERT connection

**Require:** $N_{ERT}^{DL}$,
$\quad \Omega_{ERT\_DL}^{CID} = \{CID_1 \ldots CID_i\}, \forall = 1 \ldots N_{UGS}^{DL}$

**Ensure:** $RQ_{UGS}^{DL}, RQ_{ERT}^{DL}, EQ_{ERT}^{DL}$

---

1: BEGIN
2: **for** $i = 1$ to $N_{UGS}^{DL}$ **do**
3: $\quad$ **if** $\zeta - (T_c - T_i^a(j)) \leq T_{frame}$ **then**
4: $\quad\quad P_j$ Add to $EQ_{ERT}^{DL}$
5: $\quad\quad$ /* $P_j$ : the $j$th packet for connection $i$ */
6: $\quad$ **end if**
7: **end for**
8: **for** $i = 1$ to $N_{ERT}^{DL}$ **do**
9: $\quad L_i = \sum_{j=1}^{N_j} \frac{T_i^g(j)}{N_j}, \forall = 1 \ldots N_{ERT}^{DL}, \forall = j \ldots N_j$
10: $\quad L_{max} = argmax L_i, \forall = 1 \ldots N_{ERT}^{DL}$
11: $\quad L_i = 1 - \frac{L_i}{L_{max}}, \forall = 1 \ldots N_{ERT}^{DL}$
12: $\quad$ /* $P_j$ : the $j$th packet for connection i */
13: **end for**
14: **while** $\Omega_{ERT\_DL}^{CID} > 0$ **do**
15: $\quad CID_{max} = argmin_i L_i$
16: $\quad CID_{max}$ add to $RQ_{ERT}^{DL}$
17: $\quad \Omega_{ERT\_DL}^{CID} = \Omega_{ERT\_DL}^{CID} - CID_{max}$
18: **end while**
19: $EQ_{ERT}^{DL}$ Add to $RQ_{UGS}^{DL}$
20: END

---

-RT-VR: In Algorithm 3, $N_{RT}^{DL}$ represents the number of RT-VRs in the downlink, and $\Omega_{RT\_DL}^{CID}$ represents the CID set of the RT-VRs in the downlink. The ranking is determined by the emergent degree and then sequentially pushed onto the ranking queues, $RQ_{UGS}^{DL}$.

### B.1.2 TGS

We use the symbols $RQ_{NRT}^{DL}$ and $RQ_{BE}^{DL}$ to identify the downlink ranking queues for NRT-VRs and BEs. TGS services are concerned with overall network performance, rather than packet latency. For this reason, the satisfactory degree is used as the main factor of ranking. The satisfactory degree $S_i$ is based on compensation - the fewer the number of requests served in the last frame, the higher the priority in the following frame. Here are select parameters used in the following algorithms: $T_{frame}$ represents the length of a frame, $\gamma_i$ represents the minimum reserved traffic rate of connection $i$, and $f_i(m-1)$ represents the size of the packet waiting to be served in the $(m-1)$th frame of connection $i$. $b_i^a(m-1)$ is the total bandwidth allocated from BS in the $(m-1)$th frame of connection $i$. $b_i^{NRT\_low}(m-1)$ represents the minimum bandwidth allocated from the BS to the SS in the $(m-1)$th frame of connection $I$; this bandwidth is the minimum required for the downlink to maintain a satisfactory QoS level. Equation (4) defines the minimum bandwidth request in each frame for NRT-VR connections.

$$b_i^{NRT\_low}(m-1) = min(f_i(m-1), \gamma_i T_{frame}), \forall i = 1 \ldots N_{NRT}^{DL} \quad (4)$$

Using equation (4), we can find the minimum bandwidth request for the last frame. Then, the available total bandwidth is divided by the minimum bandwidth request in last frame to find $S_i$. $S_i$

is the ratio that evaluates the satisfactory degree for connection i, as shown in equation (5).

$$S_i = \frac{b_i^a(m-1)}{b_i^{NRT\_low}(m-1)}, \forall i = 1 \ldots N_{NRT}^{DL}, \forall i \in \Omega_{NRT}^{CID} \quad (5)$$

Because there is no minimum reserved traffic rate constraint for BE, we use the number of packets waiting to be served in BS to replace it. We can calculate the satisfactory degree for BE with equation (6).

$$S_i = \frac{b_i^a(m-1)}{f_i(m-1)}, \forall i = 1 \ldots N_{BE}^{DL}, \forall i \in \Omega_{BE\_DL}^{CID} \quad (6)$$

-NRT-VR: In Algorithm 4, $N_{NRT}^{DL}$ represents the number of NRT-VRs in the downlink, and $\Omega_{NRT\_DL}^{CID}$ represents the CID set of the NRT-VRs in the downlink. The ranking is determined by the satisfactory degree and then sequentially pushed onto the ranking queues, $RQ_{NRT}^{DL}$.

---

**Algorithm 3** Priority assignment for an RT-VR connection

**Require:** $N_{RT}^{DL}$, $RQ_{RT}^{DL}$
  $\Omega_{RT\_DL}^{CID} = \{CID_1 \ldots CID_i\}, \forall = 1 \ldots N_{RT}^{DL}$
**Ensure:** $RQ_{ERT}^{DL}$, $RQ_{ERT}^{DL}$, $EQ_{RT}^{DL}$, $RQ_{RT}^{DL}$

1:   BEGIN
2:   **for** $i = 1$ to $N_{RT}^{DL}$ **do**
3:     **if** $\zeta_i - (T_c - T_i^a(j)) \leq T_{frame}$ **then**
4:       $P_j$ Add to $EQ_{RT}^{DL}$
5:       /* $P_j$ : the jth packet for connection i */
6:     **end if**
7:   **end for**
8:   **for** $i = 1$ to $N_{RT}^{DL}$ **do**
9:     $L_i = \sum_{j=1}^{N_j} \frac{T_i^a(j)}{N_j}, \forall = 1 \ldots N_{RT}^{DL}, \forall j = 1 \ldots N_j$
10:    $L_{max} = argmax L_i, \forall i = 1 \ldots N_{RT}^{DL}$
11:    $L_i = 1 - \frac{L_i}{L_{max}}, \forall i = 1 \ldots N_{RT}^{DL}$
12:    /* $P_j$ : the jth packet for connection i */
13:   **end for**
14:   **while** $\Omega_{RT_D L}^{CID} > 0$ **do**
15:     $CID_{max} = argmin_i L_i$
16:     $CID_{max}$ add to $RQ_{RT}^{DL}$
17:     $\Omega_{ERT\_DL}^{CID} = \Omega_{ERT\_DL}^{CID} - CID_{max}$
18:   **end while**
19:   $EQ_{RT}^{DL}$ Add to $RQ_{ERT}^{DL}$
20:   END

---

**Algorithm 4** Priority assignment for an NRT-VR connection

**Require:** $N_{NRT}^{DL}$
  $\Omega_{NRT\_DL}^{CID} = \{CID_1 \ldots CID_i\}, \forall i = 1 \ldots N_{NRT}^{DL}$
**Ensure:** $RQ_{NRT}^{DL}$

1:   BEGIN
2:   **for** $i = 1$ to $N_{NRT}^{DL}$ **do**
3:     $S_i = \frac{b_i^a(m-1)}{b_i^{NRT\_low}(m-1)}$
4:   **end for**
5:   **while** $\Omega_{NRT_D L}^{CID} > 0$ **do**
6:     $CID_{min} = argmin_i S_i$
7:     $CID_{min}$ add to $RQ_{NRT}^{DL}$
8:     $\Omega_{NRT\_DL}^{CID} = \Omega_{NRT\_DL}^{CID} - CID_{min}$
9:   **end while**
10:  END

---

-BE: In Algorithm 5, $N_{BE}^{DL}$ represents the number of BEs in the downlink, and $\Omega_{BE\_DL}^{CID}$ represents the CID set of the BEs in the downlink. The ranking is determined by the satisfactory degree and then sequentially pushed onto the ranking queues, $RQ_{BE}^{DL}$.

---

**Algorithm 5** Priority assignment for a BE connection

**Require:** $N_{BE}^{DL}$
  $\Omega_{BE\_DL}^{CID} = \{CID_1 \ldots CID_i\}, \forall i = 1 \ldots N_{BE}^{DL}$
**Ensure:** $RQ_{BE}^{DL}$, $EQ_{BE}^{DL}$, $RQ_{NRT}^{DL}$

1:   BEGIN
2:   **for** $i = 1$ to $N_{BE}^{DL}$ **do**
3:     **if** $\varphi_i \geq \eta$ **then**
4:       $CID_j$ Add to $EQ_{BE}^{DL}$
5:     **end if**
6:   **end for**
7:   **for** $i = 1$ to $N_{BE}^{DL}$ **do**
8:     $S_i = \frac{b_i^a(m-1)}{f_i(m-1)}$
9:   **end for**
10:  **while** $\Omega_{BE\_DL}^{CID} > 0$ **do**
11:    $CID_{min} = argmin_i S_i$
12:    $CID_{min}$ add to $RQ_{BE}^{DL}$
13:    $\Omega_{BE\_DL}^{CID} = \Omega_{BE\_DL}^{CID} - CID_{min}$
14:  **end while**
15:  $EQ_{BE}^{DL}$ Add to $RQ_{NRT}^{DL}$
16:  END

---

### B.2 Priority Elevation

In this paper, we designed a suitable priority elevation mechanism for DCS and TGS. The concept of a virtual emergent queue was proposed by us for ERT-VR and RT-VR in DCS. If the waiting time of packets exceeds the maximum latency, we will elevate the priority for services adaptively. Furthermore, if the services for ERT-VR and RT-VR connections fit equation (7), these services will be put into the emergent queue. The meaning of equation (7) is as follows: the packet can continue waiting, as long as the wait time has been less than $T_{frame}$. In fact, the so-called emergent queue inserts ERT-VR and RT-VR connections (shown in equation (13)) at the bottom of $RQ_{UGS}^{DL}$ and $RQ_{ERT}^{DL}$.

$$\zeta_i - (T_c - T_i^a(j)) \leq T_{frame} \qquad (7)$$

We utilize a service interrupt counter to observe the status of every connection in TGS and let the service interrupt connections elevate priorities to BEs. The service interrupt counter $\varphi_i$ will be used to elevate the priority of BE services. For BE services, the quality of the transmission rate is the most important factor. The service interrupt counter checks the transmission rate in the last frame. If the transmission rate is 0, $\varphi_i$ is incremented by 1. If $\varphi_i$ exceeds threshold $\eta$, the connection is presumed to be starving and has its priority elevated. That is, insert BE connections with transmission rates that exceed $\eta$ into the bottom of $RQ_{NRT}^{DL}$.

### C. Resource Allocation

As shown in Fig. 4, resource allocation is divided into two categories: bandwidth requirement quantification and bandwidth requirement allocation. Fig. 5 depicts a flowchart of resource allocation. We quantify requests to determine the allocation method. There are three cases presented in this paper with different resource allocation methods. Otherwise, we propose a weight-based proportional fairness (WPF) for TGS services to improve fairness and increase the number of served requests.

#### C.1 Bandwidth Requirement Quantification

Unsolicited grant interval, tolerated jitter, minimum reserved traffic rate, and maximum sustained traffic rate are four QoS qualifying parameters that concern DCS services. For NRT-VR services, there are two QoS parameters that need to be considered: minimum reserved traffic rate and maximum sustained traffic rate. The maximum sustained traffic rate is the main consideration for BE services.

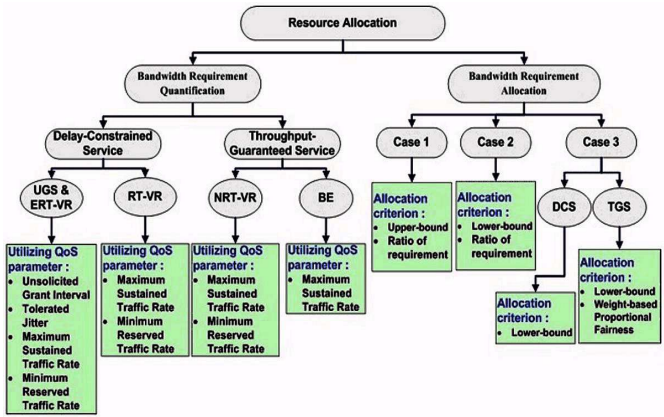

Fig. 4. Architecture of resource allocation.

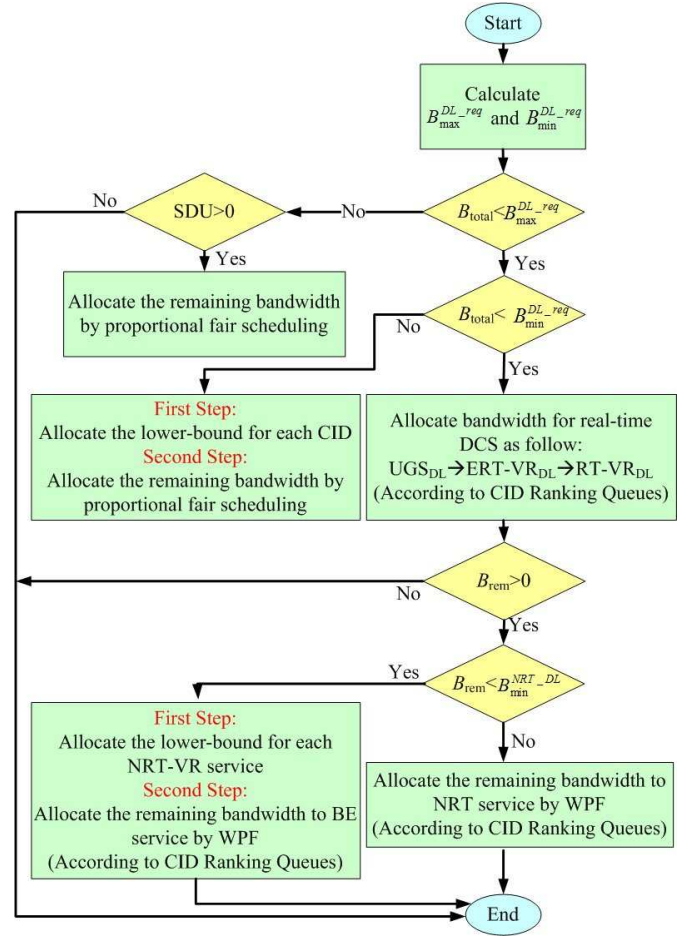

Fig. 5. Flowchart of resource allocation.

#### C.1.1 DCS

The maximum sustained traffic rate is the main factor in determining the downlink upper bound for DCS services. Similarly, the minimum reserved traffic rate is the main factor for the lower bound.

-UGS Equation (8) calculates the upper bound of UGSs for the downlink. The base station allocates bandwidth by comparing the maximum available bandwidth to the number of packets in the buffer for each connection. $b_i^{UGS\_max}(m)$ represents the maximum possible bandwidth allocation to connection $i$ in the $m$th frame, $\Gamma_i$ represents the maximum sustained traffic rate for connection $i$, and $f_i(m)$ represents the number of packets waiting to be sent to the base station.

$$b_i^{UGS\_max}(m) = min(f_i(m), \Gamma_i.T_{frame}), \forall i = 1 \ldots N_{UGS}^{DL} \qquad (8)$$

Equation (9) evaluates the sum of the upper bounds of bandwidth requests for UGSs. $b_{max}^{UGS\_DL}(m)$ represents the sum of the upper bounds of downlink bandwidth requests for UGS services.

$$b_{max}^{UGS\_DL}(m) = \sum_{i=1}^{N_{UGS}^{DL}} b_i^{UGS\_max}(m), \forall i \in \Omega_{UGS\_DL}^{CID} \qquad (9)$$

For the lower bound of bandwidth requests, the base station al-locates bandwidth by comparing the minimum requested bandwidth to the number of packets in the buffer for each connection. Equation (10) calculates the lower bound of the request. $b_i^{UGS\_min}(m)$ represents the minimum requested bandwidth of connection i in the $m$th frame, and represents the maximum sustained traffic rate for connection $i$.

$$b_i^{UGS\_min}(m) = min(f_i(m), \gamma_i.T_{frame}), \forall i = 1 \dots N_{UGS}^{DL} \quad (10)$$

Equation (11) evaluates the sum of the lower bounds of bandwidth requests for UGSs. $b_{min}^{UGS\_DL}(m)$ represents the sum of the lower bounds of bandwidth requests in the downlink for UGSs.

$$b_{min}^{UGS\_DL}(m) = \sum_{i=1}^{N_{UGS}^{DL}} b_i^{UGS\_min}(m), \forall_i \in \Omega_{UGS\_DL}^{CID} \quad (11)$$

-ERT-VR

For ERT-VR services, the upper and lower bounds are evaluated in the same way as UGSs. In equation (12), $b_i^{ERT\_max}(m)$ represents the maximum possible bandwidth allocation to connection $i$ in the $m$th frame. In equation (13), $B_{max}^{ERT\_DL}(m)$ represents the sum of the upper bounds of downlink bandwidth requests for ERT-VR services. In equation (14), $b_i^{ERT\_min}(m)$ represents the minimum requested bandwidth of connection $i$ in the $m$th frame. In equation (15),$B_{min}^{ERT\_DL}(m)$ represents the sum of the lower bounds of downlink bandwidth requests for ERT-VR services.

$$b^{ERT\_max}(m) = min(f_i(m), \Gamma_i.T_{frame}), \forall i = 1 \dots N_{ERT}^{DL} \quad (12)$$

$$B^{ERT\_max}(m) = \sum_{i=1}^{N_{ERT}^{DL}} b_i^{ERT\_max}(m), \forall i \in \Omega_{ERT\_DL}^{CID} \quad (13)$$

$$b^{ERT\_min}(m) = min(f_i(m), \gamma_i.T_{frame}), \forall i = 1 \dots N_{ERT}^{DL} \quad (14)$$

$$B^{ERT\_min}(m) = \sum_{i=1}^{N_{ERT}^{DL}} b_i^{ERT\_min}(m), \forall i \in \Omega_{ERT\_DL}^{CID} \quad (15)$$

-RT-VR

For RT-VR services, the upper and lower bounds are evaluated in the same manner as UGS and ERT-VR services. In equation (16), $b_i^{RT\_max}(m)$represents the maximum possible bandwidth allocation to connection i in the mth frame. In equation (17), $B_{max}^{RT\_DL}(m)$ represents the sum of the upper bounds of downlink bandwidth requests for RT-VR services. In equation (18), $b_i^{RT\_min}(m)$ represents the minimum requested bandwidth of connection $i$ in the $m$th frame. In equation (19), $B_{min}^{ERT\_DL}(m)$ represents the sum of the lower bounds of downlink bandwidth requests for RT-VR services.

$$b^{RT\_max}(m) = min(f_i(m), \Gamma_i.T_{frame}), \forall i = 1 \dots N_{RT}^{DL} \quad (16)$$

$$B^{RT\_max}(m) = \sum_{i=1}^{N_{RT}^{DL}} b_i^{RT\_max}(m), \forall i \in \Omega_{RT\_DL}^{CID} \quad (17)$$

$$b^{RT\_min}(m) = min(f_i(m), \gamma_i.T_{frame}), \forall i = 1 \dots N_{RT}^{DL} \quad (18)$$

$$B^{RT\_min}(m) = \sum_{i=1}^{N_{RT}^{DL}} b_i^{RT\_min}(m), \forall i \in \Omega_{RT\_DL}^{CID} \quad (19)$$

### C.1.2 TGS

In general, TGS services are concerned with the maximum sustained traffic rate and minimum reserved traffic rate for their transmission rates.

-NRT-VR

For NRT-VR services, the upper and lower bounds are evaluated in the same manner for equations (20), (21), (18), and (19). In equation (20), $b_i^{NRT\_max}(m)$ represents the maximum possible bandwidth allocation to connection $i$ in the $m$th frame. In equation (21), $B_{max}^{NRT\_DL}(m)$ represents the sum of the upper bounds of downlink bandwidth requests for NRT-VR services. In equation (22), $b_i^{NRT\_min}(m)$ represents the minimum requested bandwidth of connection $i$ in the $m$th frame. In equation (23), $B_{min}^{NRT\_DL}(m)$ represents the sum of the lower bounds of downlink bandwidth requests for RT-VR services.

$$b^{NRT\_max}(m) = min(f_i(m), \Gamma_i.T_{frame}), \forall i = 1 \dots N_{NRT}^{DL} \quad (20)$$

$$B^{NRT\_max}(m) = \sum_{i=1}^{N_{NRT}^{DL}} b_i^{RT\_max}(m), \forall_i \in \Omega_{NRT\_DL}^{CID} \quad (21)$$

$$b^{NRT\_min}(m) = min(f_i(m), \gamma_i.T_{frame}), \forall i = 1 \dots N_{NRT}^{DL} \quad (22)$$

$$B^{NRT\_min}(m) = \sum_{i=1}^{N_{ERT}^{DL}} b_i^{NRT\_min}(m), \forall i \in \Omega_{ERT\_DL}^{CID} \quad (23)$$

-BE

There is no constraint on the minimum transmission rate for BE services in downlink. Thus, we use the maximum sustained traffic rate to evaluate the upper bound of requests. In equation (24), $b_i^{BE\_max}(m)$ represents the maximum possible bandwidth allocation to connection $i$ in the $m$th frame. In equation (25), $B_{max}^{BE\_DL}(m)$ represents the sum of the upper bounds of downlink bandwidth requests for BE services.

$$b^{BE\_max}(m) = min(f_i(m), \Gamma_i.T_{frame}), \forall i = 1 \dots N_{BE}^{DL} \quad (24)$$

$$B_{max}^{BE\_DL}(m) = \sum_{i=1}^{N_{BE}^{DL}} r_i^{BE\_max}(m), \forall i \in \Omega_{BE\_DL}^{CID} \quad (25)$$

$B_{max}^{DL\_req}(m)$ represents the upper bound for total bandwidth requests, and $B_{min}^{DL\_req}(m)$ represents the lower bounds for total bandwidth requests in $m$th frame of the downlink. Because there is no lower bound constraint for BE services, we use the upper bound to replace the lower bound. According to the results of equation (26) and (27), we can choose an adaptive bandwidth requirement allocation rule dynamically.

$$B_{max}^{DL\_req}(m) = B_{max}^{UGS\_DL}(m) + B_{max}^{ERT\_DL}(m) \\ + B_{max}^{RT\_DL}(m) + B_{max}^{NRT\_DL}(m) + B_{max}^{BE\_DL}(m) \quad (26)$$

$$B_{min}^{DL\_req}(m) = B_{min}^{UGS\_DL}(m) + B_{min}^{ERT\_DL}(m) \\ + B_{min}^{RT\_DL}(m) + B_{min}^{NRT\_DL}(m) + B_{max}^{BE\_DL}(m) \quad (27)$$

### C.2 Bandwidth Requirement Allocation

In Algorithm 6, we first examine DCS services in the proposed mechanism. Then, we examine TGS services. We designed a weight-based proportional fairness (WPF) scheme for situations with insufficient remaining bandwidth for the total lower bound TGS request. To allocate bandwidth, WPF determines a weight from the number of requests and ranking. Starvation is likely to occur in TGSs. For this type of service, there is not much demand for latency. Thus, we hope to serve as many connections as possible while avoiding starvation. We utilize the WPF mechanism and set up weights ($\omega_1, \varpi_1 = 0.6$ and $\omega_2$, $\varpi_2 = 0.4$ in simulation) to increase TGS service connections, decrease starvations and maintain fairness. In Fig 4., we inspect three cases. In case I, $B_{total} > B_{max}^{DL\_req}$, we satisfy the upper bound request for all connections first. Then, we allocate the remaining bandwidth according to the ratio of unsatisfied bandwidth for connection $i$ to total unsatisfied bandwidth while maintaining the fairness principle. In case II, $B_{max}^{DL\_req} > B_{total} > B_{min}^{DL\_req}$, we still follow the fairness principle to dispatch total bandwidth resources according to the ratio of the difference in bandwidth between the upper bound request and lower bound request for an individual connection $i$ to the difference in total bandwidth request between the upper bound and lower bound. In case III, $B_{min}^{DL\_req} > B_{total}$, we satisfy the lower bound bandwidth for DCS service connections according to the queue priority first. Then, two subcases, $B_{rem} > B_{min}^{NRT}$ and $B_{rem} \leq B_{min}^{NRT}$, are considered. $B_{rem}$ represents the remaining bandwidth, and the $B_{min}^{NRT}$ represents the total lower bound request for all NRT-VR connections. If $B_{rem} > B_{min}^{NRT}$, we will satisfy the lower bound request for all NRT-VR connections first. Then, we allocate the remaining bandwidth to BE connections according to every BE connection request and priority. Otherwise, we allocate the remaining bandwidth to NRT-VR connections directly according to the bandwidth request and priority. Detailed procedures are shown for Algorithm 6.

---

**Algorithm 6** Bandwidth allocation scheme

**Require:** $B_{total}$
$\qquad RQ_{UGS}^{DL}, EQ_{BE}^{ERT}, RQ_{RT}^{DL}, RQ_{NRT}^{DL}, RQ_{BE}^{DL}$
**Ensure:** $b_i^a \forall i = 1 \dots N$

1: BEGIN
2: Calculate $B_{max}^{DL\_req}$ and $B_{min}^{DL\_req}$; $B_{rem} = B_{total}$
3: **if** $B_{total} < B_{max}^{DL\_req}$ **then**
4:    **if** $B_{total} < B_{min}^{DL\_req}$ **then**
5:      /* Allocate bandwidth for fine real-time
6:      services by Ranking Queue. The priority is as
7:      follows: $RQ_{UGS}^{DL} \rightarrow RQ_{ERT}^{DL} \rightarrow RQ_{RT}^{DL}$ */
8:      **for** $i = 1$ to $N_{fine-real-time}$ **do**
9:        **if** $B_{rem} > 0$ **then**
10:          $b_i^a = b_i^{min}$
11:          $B_{rem} = B_{rem} - b_i^{min}$
12:        **end if**
13:      **end for**
14:      **if** $B_{rem} > 0$ **then**
15:        **if** $B_{rem} > B_{min}^{NRT}$ **then**
16:          **for** $i = 1$ to $N_{NRT}$ **do**
17:            $b_i^a = b_i^{min}$
18:            $B_{rem} = B_{rem} - b_i^{min}$
19:          **end for**
20:          /* Make use of WPF scheme to allocate the
21:          remaining bandwidth for BE services by $RQ_{BE}^{DL}$*/
22:          **for** $i = 1$ to $N_{BE}$ **do**
23:            $b_i^a = [\frac{b_i^{max}}{B_{max}^{BE\_DL}} * \omega_1 + \frac{\varphi_i}{\sum_i^{N_{BE}} \varphi_i} * \omega_2] * B_{rem}$
24:          **end for**
25:        **else**
26:          /* Make use of WPF scheme to allocate the
27:          remaining bandwidth for NRT services by $RQ_{NRT}^{DL}$ */
28:          **for** $i = 1$ to $N_{NRT}$ **do**
29:            $b_i^a = [\frac{b_i^{max}}{B_{max}^{NRT\_DL}} * \varpi_1 + \frac{\varphi_i}{\sum_i^{N_{BE}} \varphi_i} * \varpi_2] * B_{rem}$
30:          **end for**
31:        **end if**
32:      **end if**
33:    **else**
34:      **for** $i = 1$ to $N$ **do**
35:        $b_i^a = b_i^{min}$
36:        $B_{rem} = B_{rem} - b_i^{min}$
37:      **end for**
38:      **if** $B_{rem} > 0$ **then**
39:        **for** $i = 1$ to $N$ **do**
40:          $b_i^a = b_i^{min} + \frac{b_i^{max} - b_i^{min}}{B_{max} - B_{min}} * B_{rem}$
41:        **end for**
42:      **end if**
43:    **end if**
44: **else**
45:    **for** $i = 1$ to $N$ **do**
46:      $b_i^a = b_i^{max}$
47:      $B_{rem} = B_{rem} - b_i^{max}$
48:    **end for**
49:    **for** $i = 1 to N$ **do**
50:      **if** $SDU - b_i^{max} > 0$ **then**
51:        $b_i^a = b_i^{min} + \frac{SDU_i - b_i^{max}}{\sum_i^N SDU_i - b_i^{max}} * B_{rem}$
52:      **end if**
53:    **end for**

## III. SIMULATION RESULTS

### A. Environment and Parameters

#### A.1 Scenario

The simulation used a point-to-multipoint network architecture that comprised one base station (BS) and nine mobile stations (MS), as shown in Fig 6. Table 2 shows the service type and CID of each mobile station.

#### A.2 Assumptions

-A TDD-based model was used.

-Scheduling was decided by the BS, taking into consideration the downlink.

-In the BS and MS, packets were dropped if the queue was full.

-All connections were set up after call admission control.

-Connections were not made or canceled during the simulation.

#### A.3 Parameters

Table 3 shows the simulation parameters, as well as the WPF weight setting (the sum of all weights is 1) and the service interrupt counter threshold [38-40]. The total number of connections was 45. We defined the queue size and packet size for different types of services. The total amount of bandwidth was 10 Mbps, and the frame duration was 5 ms. Simulation time was 10 sec (2000 frames). The service interrupt counter threshold $\eta$ was 50.

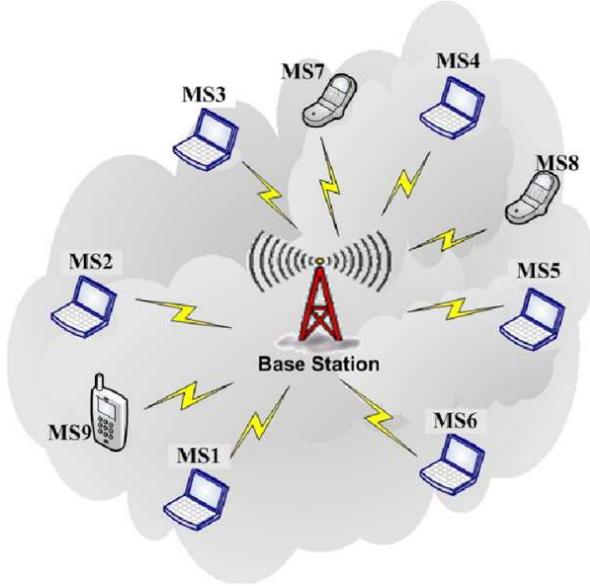

Fig. 6. Scenario architecture.

Table 2. THE SERVICE TYPE AND CID OF EACH MOBILE STATION

| MS1 | | MS2 | | MS3 | |
|---|---|---|---|---|---|
| CID Number | Services Type | CID Numbe | Services Type | CID Numbe | Services Type |
| CID1 | UGS-DL | CID6 | UGS-DL | CID11 | UGS-DL |
| CID2 | ERT-VR-DL | CID7 | ERT-VR-DL | CID12 | ERT-VR-DL |
| CID3 | RT-VR-DL | CID8 | RT-VR-DL | CID13 | RT-VR-DL |
| CID4 | NRT-VR-DL | CID9 | NRT-VR-DL | CID14 | NRT-VR-DL |
| CID5 | BE-DL | CID10 | BE-DL | CID15 | BE-DL |
| MS4 | | MS5 | | MS6 | |
| CID Number | Service Type | CID Numbe | Service Type | CID Numbe | Service Type |
| CID16 | UGS-DL | CID21 | UGS-DL | CID26 | UGS-DL |
| CID17 | ERT-VR-DL | CID22 | ERT-VR-DL | CID27 | ERT-VR-DL |
| CID18 | RT-VR-DL | CID23 | RT-VR-DL | CID28 | RT-VR-DL |
| CID19 | NRT-VR-DL | CID24 | NRT-VR-DL | CID29 | NRT-VR-DL |
| CID20 | BE-DL | CID25 | BE-DL | CID30 | BE-DL |
| MS7 | | MS8 | | MS9 | |
| CID Number | Service Type | CID Numbe | Service Type | CID Numbe | Service Type |
| CID31 | UGS-DL | CID36 | UGS-DL | CID41 | UGS-DL |
| CID32 | ERT-VR-DL | CID37 | ERT-VR-DL | CID42 | ERT-VR-DL |
| CID33 | RT-VR-DL | CID38 | RT-VR-DL | CID43 | RT-VR-DL |
| CID34 | NRT-VR-DL | CID39 | NRT-VR-DL | CID44 | NRT-VR-DL |
| CID35 | BE-DL | CID40 | BE-DL | CID45 | BE-DL |

Table 3. QOS PARAMETERS

| Parameter | Value |
|---|---|
| Number of Connections | 45 |
| Number of MSs | 9 |
| Queue Size | 100 |
| Packet size of UGS | 160 Byte |
| Packet size of ERT-VR | 160 Byte |
| Packet size of RT-VR | 240 Byte |
| Packet size of NRT-VR | 120 Byte |
| Packet size of BE | 120 Byte |
| Total amount of bandwidth | 10 Mbps |
| Frame duration | 5 ms |
| Simulation time | 10 sec (2000 frames) |
| Service Interrupt Counter ( $\eta$ ) | 50 (250 ms) |
| $\omega_1$ | 0.6 |
| $\omega_2$ | 0.4 |
| $\varpi_1$ | 0.6 |
| $\varpi_2$ | 0.4 |

### B. Simulation Results and Analysis

We will now subject APDS to average delay and average throughput comparisons with other related standard scheduling schemes: first in first out (FIFO), deficit fair priority queue (DFPQ) [16], and single-carrier scheduling algorithm (SCSA) [20, 21].

#### B.1 UGS

In Fig. 7, we can see that the average throughput of APDS is better than the other methods for efficient scheduling. The average delay of UGSs in the downlink is shown in Fig. 8. Because APDS considers the average delay as a main factor in its scheduling algorithm, APDS has shorter average delays compared to the other methods.

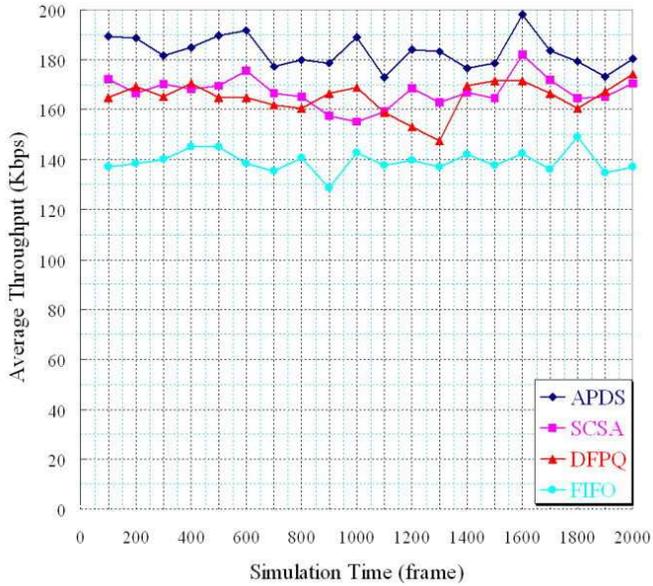

Fig. 7.  Average Throughput of UGS (Downlink).

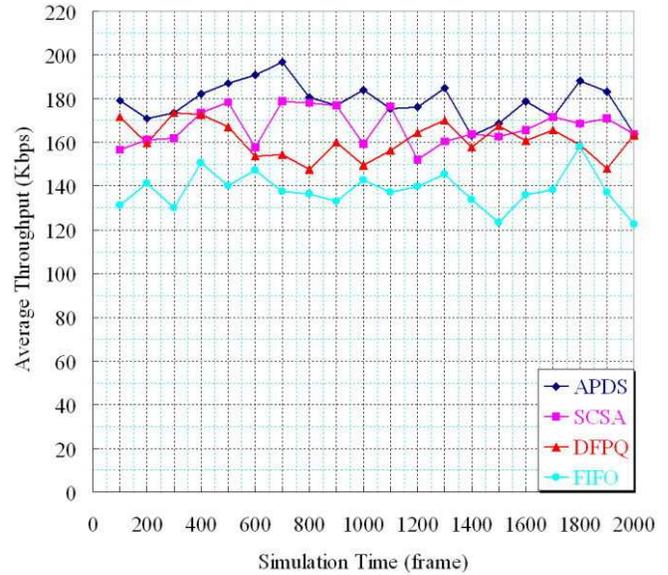

Fig. 9.  Average Throughput of ERT-VR (Downlink).

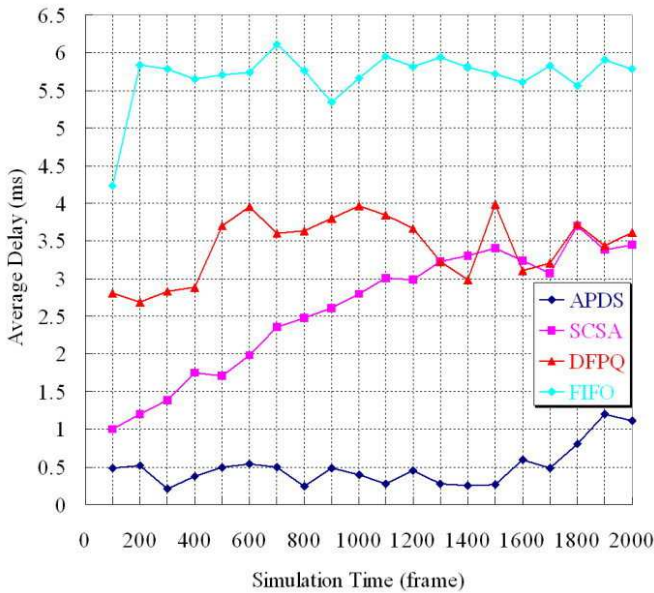

Fig. 8.  Average Delay of UGS (Downlink).

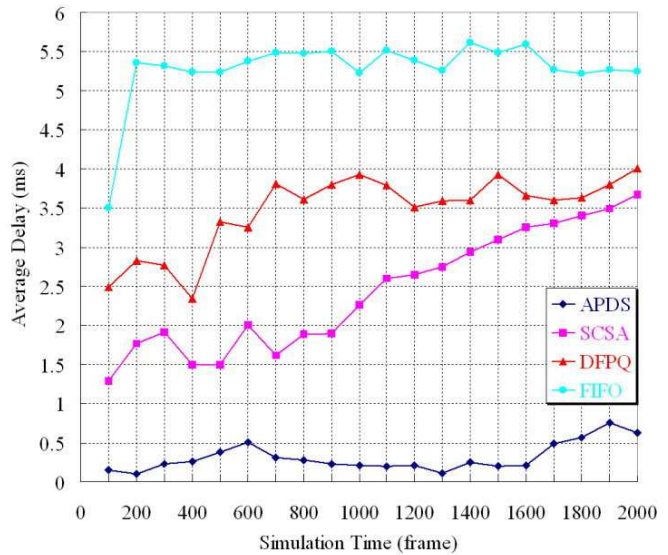

Fig. 10.  Average Delay of ERT-VR (Downlink).

### B.2  ERT-VR

Fig. 9 and 10 show the average throughput and average delay, respectively. APDS considers the average delay as a main factor in its scheduling algorithm and utilizes an emergent queue to increase the emergent packet transfer probability. As shown in the results, the performance of APDS is better than the other methods.

### B.3  RT-VR

Fig. 11 and 12 show the average throughput and average delay for the downlink, respectively. APDS considers the average delay as a main factor in its scheduling algorithm and utilizes an emergent queue to increase the emergent packet transfer probability. As shown in the results, the performance of APDS is better than the other methods.

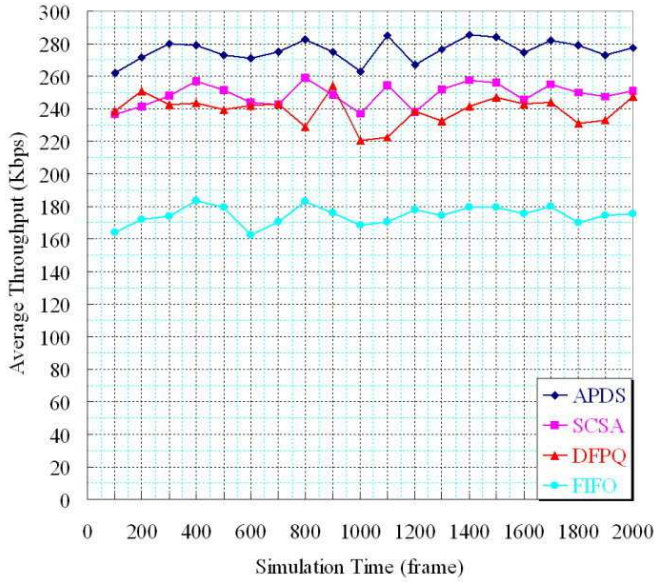

Fig. 11.  Average Throughput of RT-VR (Downlink).

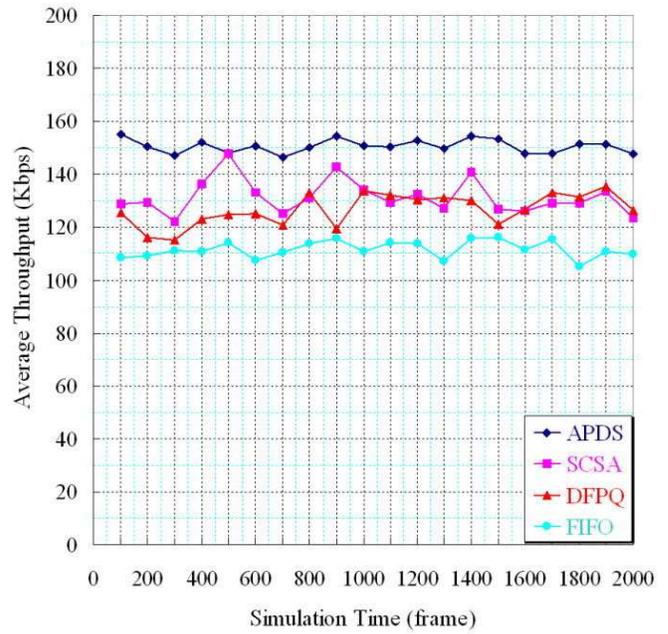

Fig. 13.  Average Throughput of NRT-VR (Downlink).

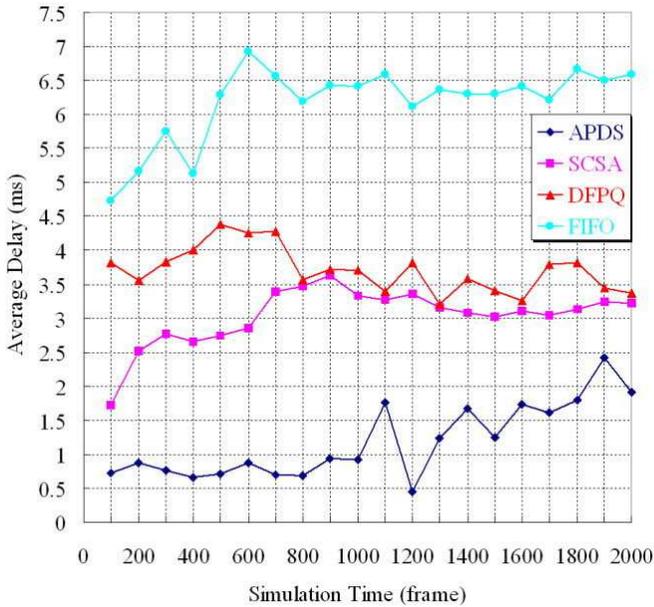

Fig. 12.  Average Delay of RT-VR (Downlink).

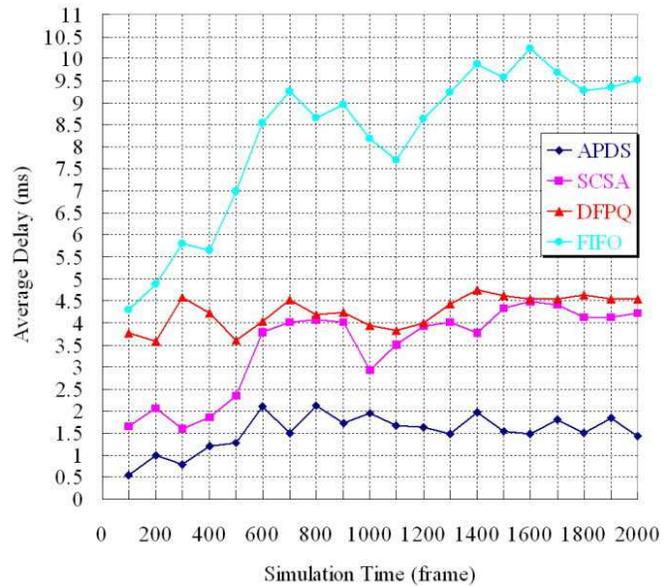

Fig. 14.  Average Delay of NRT-VR (Downlink).

### B.4  NRT-VR

Fig. 13 and 14 show the average throughput and average delay in the downlink, respectively.  For the NRT-VR services, APDS uses a performance provision for ranking to increase network performance and utilizes WPS in resource allocation to decrease the probability of service interrupts.  As shown in the results, the performance of APDS is better than the other methods.

### B.5  BE

Fig. 15 and 16 show the average throughput and average delay in the downlink, respectively.  APDS uses a performance provision for ranking to increase network performance and utilizes a service interrupt counter to avoid BE service interrupts. For resource allocation, WPF increases the priority of services to avoid interrupts. As shown in the results, the performance of APDS is better than FIFO and SCSA. The round-robin method is used for fairness in DFPQ. For this reason, DFPQ has better performance than APDS in the downlink for BE services.

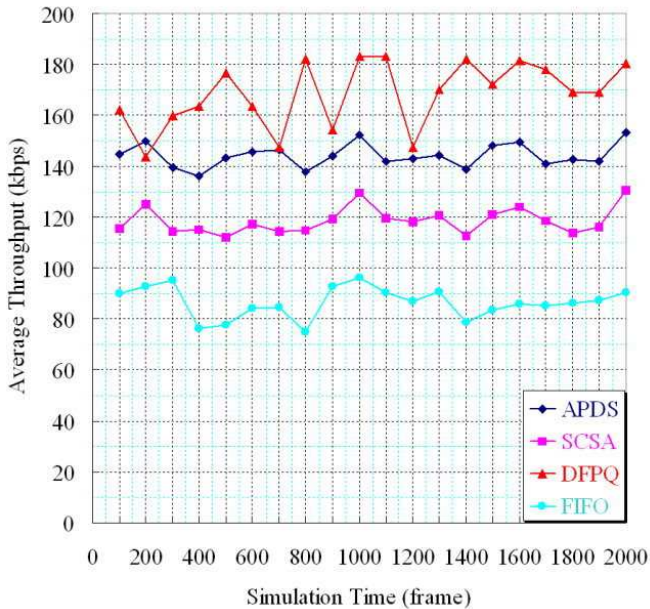

Fig. 15.  Average Throughput of BE (Downlink).

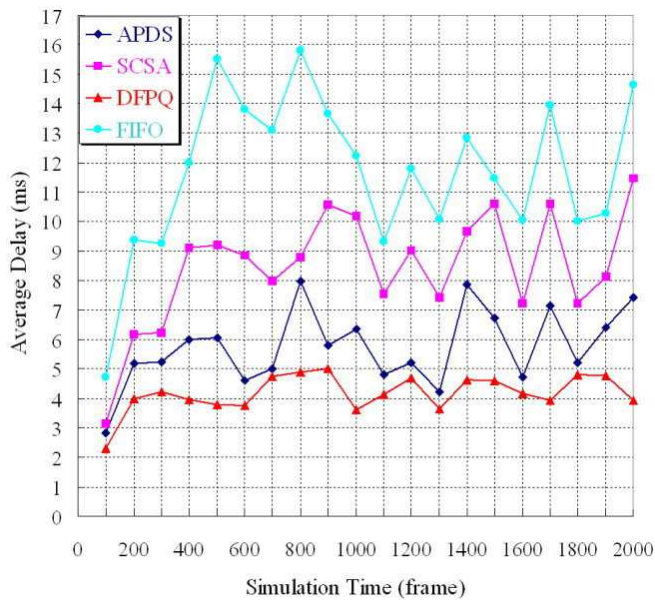

Fig. 16.  Average Delay of BE (Downlink).

## IV.  CONCLUSIONS

This paper proposes an adaptive priority-based downlink scheduling framework for multilevel downlink traffic in IEEE 802.16 networks. Our APDS framework introduces beneficial schemes to not only rank the connections of the separate service types based on the determined priority, but also to achieve QoS guarantees and starvation prevention. Additionally, the proposed bandwidth allocation scheme is well designed for QoS differentiation and satisfaction. The simulation results reveal that APDS has significant performance advantages over FIFO, DFPQ, and SCSA. We will extend this work to the uplink and consider IEEE 802.16j in the future.

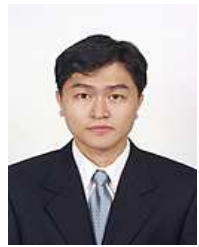

**Shih-Jung Wu** was born in Taipei, Taiwan (R.O.C.), on October 25, 1976. He received his B.C. degree from Department of Business Administration, Yuan Ze University, Taiwan (R.O.C.) in 1998. He received M.S. degree from Department of Computer Science and Information Engineering, Tamkang University, Taiwan (R.O.C.) in 2001. And he received his Ph.D. degree in the Department of Computer Science and Information Engineering, Tamkang University, Taiwan (R.O.C.) in 2006. Presently, he is working at Department of Innovative Information and Technology, Tamkang University in Taiwan (R.O.C.). His major research interests in high speed communications, mobility, QoS guarantees, parallel algorithms and data mining.

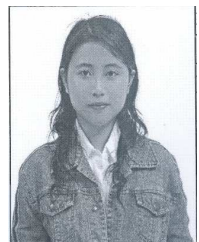

**Shih-Yi Huang** She received M.S. degree from Master's Program in Networking and Communications, Department of Computer Science and Information Engineering, Tamkang University, Taiwan (R.O.C.) in 2008. Presently, she is working at Chunghwa Telecom Laboratories.

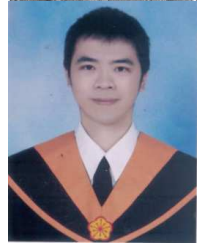

**GKuo-Feng Huang** He received M.S. degree from Department of Computer Science and Information Engineering, Tamkang University, Taiwan (R.O.C.) in 2007. And he received his Ph.D. degree in the Department of Computer Science and Information Engineering, Tamkang University, Taiwan (R.O.C.) in 2011. Presently, he is working at Department of Information Technology and Mobile Communication, Taipei College of Maritime Technology. His major research interests in wireless network.

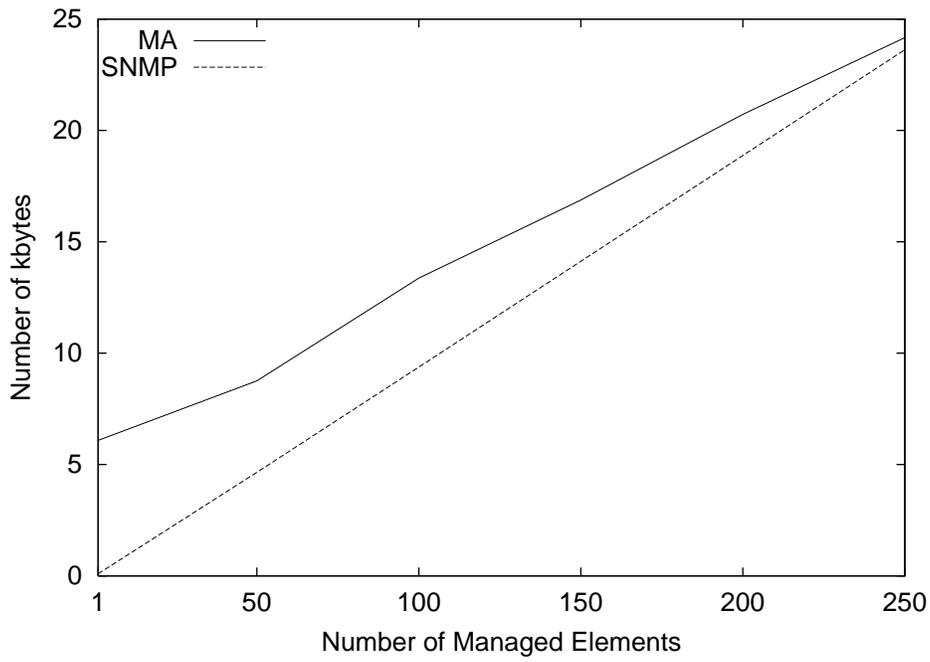

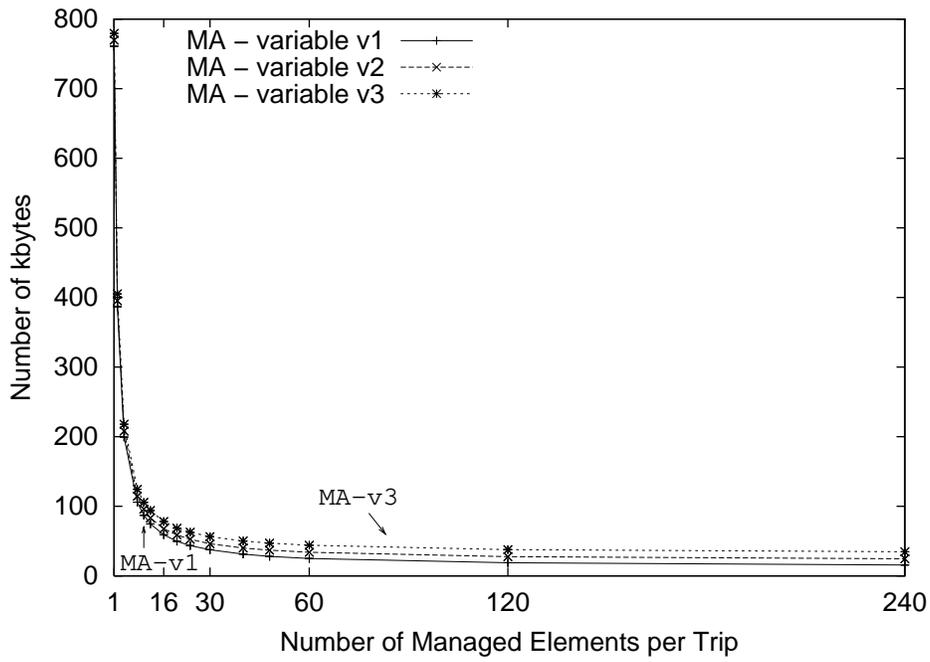

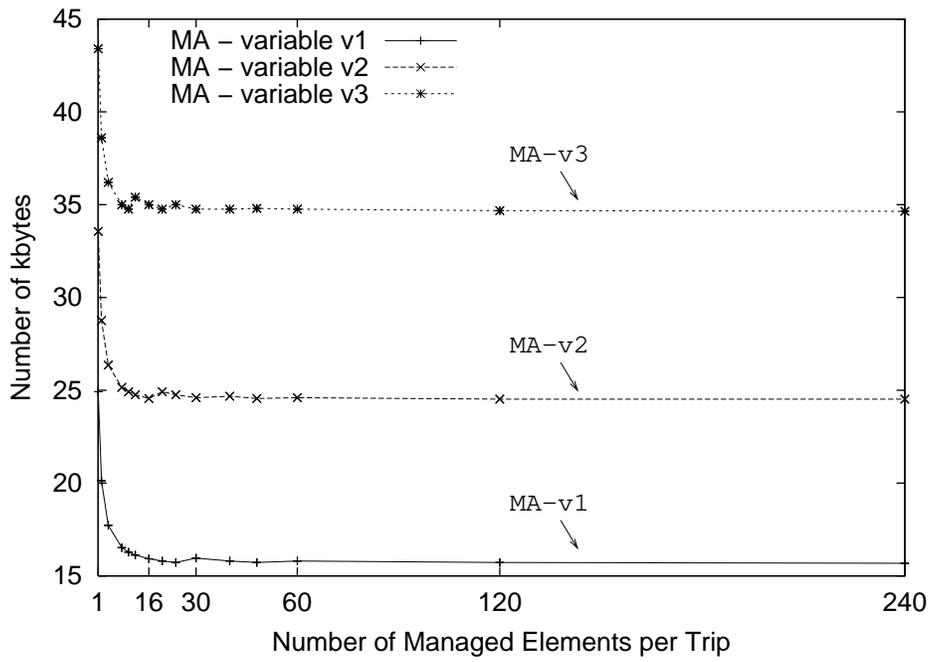